\documentstyle[12pt,epsf,epsfig]{article}
\def\beq{\begin{equation}}
\def\eeq{\end{equation}}
\def\bea{\begin{eqnarray}}
\def\eea{\end{eqnarray}}
\def\bq{\begin{quote}}
\def\eq{\end{quote}}

\def\gappeq{\mathrel{\rlap {\raise.5ex\hbox{$>$}}
{\lower.5ex\hbox{$\sim$}}}}

\def\lappeq{\mathrel{\rlap{\raise.5ex\hbox{$<$}}
{\lower.5ex\hbox{$\sim$}}}}

\def\tolim{\mathop{\longrightarrow}\limits} 
\def\simeq{\buildrel\sim\over =}

\begin{document}

\renewcommand{\thesection}{\arabic{section}.}
\renewcommand{\theequation}{\arabic{section}.\arabic{equation}}
\pagestyle{empty}
\begin{flushright}
CERN-TH/98-129\\
RU98-3-B\\
LAPTH683/98\\
LPTHE ORSAY 98/31
\end{flushright}

\begin{center}
{\bf UNIVERSALITY OF LOW-ENERGY SCATTERING IN (2+1) DIMENSIONS}\\
\vspace{0.3cm}

{\bf Khosrow Chadan}\\
Laboratoire de Physique Th\'eorique et Hautes Energies\footnote{Laboratoire
associ\'e au CNRS URA 63.}\\
Universit\'e Paris-Sud, F-91405 Orsay, France\\
\vspace{0.1cm}  

{\bf N. N. Khuri}\\
Department of Physics\\
The Rockefeller University, New York, New York 10021\\
\vspace{0.1cm}  

{\bf Andr\'e Martin}\\
TH Division, CERN, Geneva, Switzerland,\\
Laboratoire de Physique Th\'eorique ENSLAPP\\
F-74941 Annecy-le-Vieux, France, and\\
Department of Physics, The Rockefeller University\\
New York, New York 10021\\
\vspace{0.1cm}  

{\bf Tai Tsun Wu}\\
Gordon McKay Laboratory, Harvard University\\
Cambridge, Massachusetts 02138-2901, and\\
TH Division, CERN, CH - 1211 Geneva 23\\

\vspace{0.3cm}  
{\bf ABSTRACT}\\

\end{center}

\indent 
We prove that, in (2+1) dimensions, the $S$-wave phase
shift, $\,\delta_0(k)$, $\,k$ being the c.m.\\ momentum, vanishes as either
\begin{displaymath}
\delta_0\to {c\over \ln (k/m)}\quad\mbox{or}\quad \delta_0\to O(k^2)
\end{displaymath}
as $k\to 0$.  The constant $c$ is universal and $c=\pi/2$.  This
result is established first in the framework of the Schr\"odinger equation
for a large class of potentials, second for a massive field theory from proved
analyticity and unitarity, and, finally, we look at perturbation theory in
$\phi_3^4$ and study its relation to our non-perturbative result.  The
remarkable fact here is that in $n$-th order the perturbative amplitude diverges
like $(\ln k)^n$ as
$k\to 0$, while the full amplitude vanishes as $(\ln k)^{-1}$.  We show
how these two facts can be reconciled. 
\begin{flushleft}
CERN-TH/98-129\\
RU98-3-B\\
LAPTH683/98\\
LPTHE ORSAY 98/31 \\
April 1998
\end{flushleft}

\eject
\pagestyle{plain}
\setcounter{page}{1}
\section{INTRODUCTION}

Quantum field theories in $2+1$ dimensions provide us with a useful field
of investigation not only for theoretical and mathematical issues, but
also in some cases for actual physical problems. 
Scattering in 2+1 dimensions has the advantage of resembling scattering
in 3+1 dimensions much more than 1+1 dimensions in the sense that the
scattering amplitude depends on two variables and that, also, the
scattering matrix is non-trivial $\underline{\rm only}$ if production
processes exist. As we shall see in Section 3, the analyticity-unitarity
programme follows the same lines as in 3+1 dimensions. On the other hand,
a non-trivial $\lambda\phi^4$ massive field  theory can be constructed in
2+1 dimensions \cite{one}. Another merit of two space
dimensions is that it covers situations occurring in condensed matter
physics.

A remarkable property, which is completely different from what happens in
3+1 dimensions is the threshold behaviour of the scattering amplitude. 
If $\delta_0$ is the $S$ wave phase shift, then, as $k\rightarrow 0$
either\
\beq
\delta_0 \sim {\pi\over 2\ln k}
\label{oneone}
\eeq
or
\beq
\delta_0 = 0(k^2)
\label{onetwo}
\eeq

In this paper we prove this result in three situations:
\begin{itemize}
\item[i)] in the framework of the Schr\"odinger equation for a very large
class of potentials under conditions which will be specified in Section 2
and which are presumably very difficult to weaken; in this case, it
appears that the generic case is given by Eq. (\ref{oneone}), Eq.
(\ref{onetwo}) representing an exceptional case; this is done in Section 2;
\item[ii)] in massive axiomatic field theory, in combination with
unitarity and a standard assumption of smooth behaviour of the scattering
amplitude; this is carried in Section 3;
\item[iii)]  starting from the perturbation
expansion of $\lambda\phi^4$ and resumming the leading logs, while
individual terms diverge like (ln $k)^n$ near $k = 0$. 
This is done in
Section 4. 
\end{itemize}
To us and to many of our friends and colleagues, property
(\ref{oneone})-(\ref{onetwo}) was a great surprise even though a related property appeared
already in the theory of antennas \cite{three}.

However, it turned out that we had predecessors. Concerning i), the
potential case, Boll\'e and Gestezy \cite{four} found property (\ref{oneone}) for a
class of potentials exponentially decreasing at infinity and Averbush
\cite{five} treated the case of a potential with a strictly finite range.
However,  these authors missed the exceptional case (\ref{onetwo}).

Concerning ii) and iii), Bros and Iagolnitzer \cite{six}, in the
framework of general $S$ matrix theory, studying primarily the Riemann
sheet structure of scattering amplitudes near threshold, obtained for the
two-body $\rightarrow$ two-body case an equation which implies the
alternative between (\ref{oneone}) and (\ref{onetwo}). If one combines 
their paper with the
early work of Bros, Epstein and Glaser on analyticity of the scattering
amplitude near physical points \cite{seven} one can consider this as an
axiomatic proof.  They also carry the resummation of a subclass of
perturbation graphs, without, however, showing that they are the ones
associated to the leading logs.

In spite of this, we believe that we have the duty to present our own
results with our own methods, which we believe sometimes more accessible,
covering in a synthetic way the potential case and the field theoretical
case and, in some instances, improving or correcting previous work.

\section{TWO-DIMENSIONAL POTENTIAL SCATTERING}
\setcounter{equation}{0}

In two dimensions the partial wave expansion of the scattering amplitude
$T(k,\theta)$ is given by
\begin{equation}
T(k,\theta)={1\over\sqrt{k}}\,\sum_{n=0}^{\infty}\,\epsilon_n\,
(e^{i\delta_n}\sin\delta_n)\,\cos n\theta,\label{eq:2.1}
\end{equation}
where $\epsilon_0=1$, $\,\epsilon_n=2$ for $n\ge 1$.  The phase shifts
$\delta_n(k)$ are obtained in the standard way from the solutions of the
Schr\"odinger equation.  In this paper we are interested mainly in the
term $n=0$.

The $n=0$ solutions, $\,u(k,r)$, satisfy
\begin{equation}
\biggl[{d^2\over dr^2}+{1\over
4r^2}+k^2-gV(r)\biggr]\,u(k,r)=0.\label{eq:2.2}
\end{equation}
Without loss of generality, $\,g$ is taken to be non-negative.
Equation (\ref{eq:2.2}), under conditions on $V(r)$ to be specified below,
has two independent solutions: behaving like $\sqrt{r}$ and $\sqrt{r}\,\ln
r$ as $r\to 0$.  We take as a regular solution
\begin{equation}
u(k,0)=0,\quad u(k,r)\sim \sqrt{r},\label{eq:2.3}
\end{equation}
corresponding to a finite wave function at the origin.  For a discussion
of this choice, see Appendix~A.

The phase shift, $\,\delta_0(k)$, is defined by
\begin{equation}
u(k,r)\tolim_{r\to\infty} c\sqrt{r}\,[\cos
\delta_0\,J_0(kr)-\sin\delta_0\,Y_0(kr)].\label{eq:2.4}
\end{equation}
The sign of the second term is chosen to correspond to the definition of
$\delta_0$ in the three-dimensional case, i.e., $\,u\to
c\sqrt{\pi/2}\,\cos(kr-\pi/4 +\delta_0)$ as $r\to
\infty$.  By rearranging terms in Eq.~(\ref{eq:2.4}), we get 
\begin{equation}
u(k,r)\tolim_{r\to\infty} c\sqrt{r}\,e^{-i\delta_0}\,[H_0^{(2)}(kr)
+e^{2i\delta_0}H_0^{(1)}(kr)].\label{eq:2.5}
\end{equation}
We can always choose $u(k,r)$ such that
\begin{equation}
u(k,r)\tolim_{r\to\infty} -{1\over 2i\sqrt{k}}\,[e^{-i(kr-\pi/4)}
+S(k)e^{+i(kr-\pi/4)}],\label{eq:2.6}
\end{equation}
where we have used the asymptotic formulas for $H_0^{(1),(2)}(z)$ for
large $|z|$, and
\begin{equation}
S(k)\equiv e^{2i\delta_0(k)}.\label{eq:2.7}
\end{equation}

The Jost functions in this case are solutions of (\ref{eq:2.2}) finite at
$r=0$, which we denote as $f_{\pm}(k,r)$ with the asymptotic behaviour
\begin{equation}
f_{\pm}(k,r)\tolim_{r\to\infty} e^{\mp i(kr-\pi/4)}.\label{eq:2.8}
\end{equation}
We can thus write
\begin{equation}
u(k,r)=-{1\over 2i\sqrt{k}}\,[f_{+}(k,r)+S(k)f_{-}(k,r)].\label{eq:2.9}
\end{equation}

It is convenient to follow a method of treating singular
potentials \cite{ten}.We shall see below how this simplifies the
task of taking the limit $k\to 0$.  Following ref. \cite{ten}, we define $g(k,r)$ as
\begin{equation}
g(k,r)\equiv {1\over 2i\sqrt{k}}\,[f_{+}(k,r)+f_{-}(k,r)].\label{eq:2.10}
\end{equation}
The sign here is different from that in the three-dimensional case.  From Eq.\
(\ref{eq:2.9}) we now have
\begin{equation}
u(k,r)=-[g(k,r) +A(k)f_{-}(k,r)],\label{eq:2.11}
\end{equation}
where $A(k)$ is the $n=0$ scattering amplitude,
\begin{equation}
A(k)\equiv {1\over 2i\sqrt{k}}\,[S(k)-1]\equiv
{1\over\sqrt{k}}\,e^{i\delta_0}\sin\delta_0.\label{eq:2.12}
\end{equation}
The condition $u(k,r)\to 0$ as $r\to 0$ gives us
\begin{equation}
A(k)=-\lim_{r\to0}\,[g(k,r)/f_{-}(k,r)].\label{eq:2.13}
\end{equation}
Notice that this limit is always finite.  This is because $f_{-}$, being a
combination of ${\rm Re}\ f_{-}$ and ${\rm Im}\ f_{-}$, i.e., of two
linearly independent solutions of (\ref{eq:2.2}), has to behave as
$f_{-}\sim\sqrt{r}\,\ln r$ as $r\to0$.
 
The asymptotic behaviour of $u(k,r)$ can be written as
\begin{equation}
u(k,r)\tolim_{r\to\infty} {i\cos (kr-\pi/4)\over\sqrt{k}}
-A(k)e^{i(kr-\pi/4)}.\label{eq:2.14}
\end{equation}
This follows from Eqs.\ (\ref{eq:2.8}) and (\ref{eq:2.9}).

Following ref. \cite{ten}, we introduce a Green's function $G(r,r')$ for $r,r'>0$,
defined by
\begin{equation}
\biggl[{d^2\over dr^2}+{1\over 4r^2}+k^2\biggr]G(r,r')\equiv
\delta(r-r').\label{eq:2.15}
\end{equation}
This $G$ is given explicitly by
\begin{equation}
G(r,r')={\pi\over 2}\,\sqrt{rr'}\,[J_0(kr)Y_0(kr')-J_0(kr')Y_0(kr)]
\theta(r'-r),\label{eq:2.16}
\end{equation}
where $J_0$ and $Y_0$ are the standard Bessel functions of the first and
second kind.

The next step is to introduce a $u_0(k,r)$ which is a solution of the
free, $\,V=0$, Schr\"odinger equation.  We set
\begin{equation}
u_0(k,r)\equiv u(k,r)- g\int_0^{\infty}
dr'\,G(r,r')V(r')u(k,r').\label{eq:2.17}
\end{equation}
From Eq.\ (\ref{eq:2.15}) it is now obvious that
\begin{equation}
\biggl[{d^2\over dr^2}+{1\over 4r^2}+k^2\biggr]u_0(k,r)=0.\label{eq:2.18}
\end{equation}
As $r\to\infty$, $\,u_0\to u$, and from Eq.\ (\ref{eq:2.14}) it is clear
that $u_0$ is given by
\begin{equation}
u_0(k,r)=\sqrt{{\displaystyle {\pi\over 2}}}\,i\sqrt{r}\,J_0(kr) - 
\sqrt{{\displaystyle {\pi\over
2}}}\,A(k)\sqrt{kr}\,H_0^{(1)}(kr).\label{eq:2.19}
\end{equation}

The integral equation for $u$ can now be written as
\begin{equation}
u(k,r)=u_0(k,r)+g\int_r^{\infty}dr'\,\tilde G
(k;r,r')V(r')u(k,r'),\label{eq:2.20}
\end{equation}
with
\begin{equation}
\tilde G (k;r,r')={\pi\over 2}\,\sqrt{rr'}\,[J_0(kr)Y_0(kr')
-J_0(kr')Y_0(kr)].\label{eq:2.21}
\end{equation}
Using Eqs.\ (\ref{eq:2.11}) and (\ref{eq:2.19}), we can get from
(\ref{eq:2.20}) two separate integral equations for $g(k,r)$ and
$f_{-}(k,r)$.  These are
\begin{equation}
g(k,r)=-i\sqrt{{\displaystyle {\pi\over 2}}}\,\sqrt{r}\,J_0(kr)
+g\int_r^{\infty}dr'\,\tilde G(k;r,r')V(r')g(k,r')\label{eq:2.22}
\end{equation}
and
\begin{equation}
f_{-}(k,r)=\sqrt{{\displaystyle {\pi\over 2}}}\,\sqrt{kr}\,H_0^{(1)}(kr)
+g\int_r^{\infty} dr'\,\tilde G(k;r,r')V(r')f_{-}(k,r').\label{eq:2.23}
\end{equation}
These last two equations are the same except for the inhomogeneous term. 
We are interested in studying them in the limit of small~$k$.  Before we
can do that, it is convenient to remove a $\sqrt{k}$ factor from
$f_{-}$ and define $\tilde f_{-}(k,r)$ as
\begin{equation}
\tilde f_{-}(k,r)\equiv {1\over\sqrt{k}}\,f_{-}(k,r).\label{eq:2.24}
\end{equation}
With this definition, Eq.\ (\ref{eq:2.13}) becomes
\begin{equation}
e^{i\delta_0(k)}\sin\delta_0(k)=-\lim_{r\to 0}\,[g(k,r)/\tilde
f_{-}(k,r)].\label{eq:2.25}
\end{equation}

We now take the $k\to0$ limit of Eq.\ (\ref{eq:2.22}) and the equation
corresponding to (\ref{eq:2.23}) for $\tilde f_{-}$.  Using
\begin{equation}
{\pi\over 2}\,[J_0(kr)Y_0(kr')-J_0(kr')Y_0(kr)] =\ln {r'\over
r} +O(k^2)\label{eq:2.26}
\end{equation}
for small~$k$, we get
\begin{equation}
g(k,r)=-i\sqrt{{\displaystyle {\pi\over 2}}}\,\sqrt{r}
+g\int_r^{\infty} dr'\,\sqrt{rr'}\,\biggl(\ln {r'\over
r}\biggr)V(r')g(k,r')+O(k^2)\label{eq:2.27}
\end{equation}
and
\begin{eqnarray}
\tilde f_{-}(k,r)&=&i\sqrt{{\displaystyle {2\over \pi}}}\,\biggl(\ln k +\ln
r-\ln 2 +\gamma -i\,{\pi\over 2}\biggr)\,\sqrt{r}\nonumber\\
&&\mbox{} +g\int_r^{\infty} dr'\,\,\sqrt{rr'}\,\biggl(\ln {r'\over
r}\biggr)V(r')\tilde f_{-}(k,r')+O(k^2),\label{eq:2.28}
\end{eqnarray}
where $\gamma$ is Euler's constant.  For $r>0$, taking the $k\to 0$ limit
under the integral sign is allowed if we assume
\begin{equation}
\int_a^{\infty}r'\,dr'\,(1+|\ln r'|^2)\,|V(r')|<\infty,\quad
a>0.\label{eq:2.29}
\end{equation}
We shall discuss this condition in more detail later.

At this stage, we introduce two functions, $\,A(r)$ and $B(r)$, defined
by the following integral equations:
\begin{equation}
A(r)=1+g\int_r^{\infty}r'\,dr'\,\biggl(\ln {r'\over
r}\biggr)V(r')A(r')\label{eq:2.30}
\end{equation}
and
\begin{equation}
B(r)=\ln r +g\int_r^{\infty}r'\,dr'\,\biggl(\ln {r'\over r}\biggr)
V(r')B(r').\label{eq:2.31}
\end{equation}
It is clear from inspecting Eqs.\ (\ref{eq:2.27}) and (\ref{eq:2.28}) that
\begin{equation}
A(r)\equiv \lim_{k\to 0}\left[{ig(k,r)\over
\sqrt{\pi/2}\,\sqrt{r}}\right]\label{eq:2.32}
\end{equation}
and
\begin{equation}
\left[{-i\tilde f_{-}(k,r)\over \sqrt{2/\pi}\,\sqrt{r}}\right]\equiv
\biggl[A(r)\biggl(\ln k-\ln 2+\gamma -i\,{\pi\over
2}\biggr)+B(r)\biggr]+O(k^2).\label{eq:2.33}
\end{equation}
Thus, for small $k$ we have
\begin{equation}
-\biggl[{g(k,r)\over \tilde f_{-}(k,r)}\biggr]={(\pi/2)A(r)\over
{\displaystyle A(r)\biggl(\ln k-\ln 2+\gamma-i\,{\pi\over 2}\biggr) +B(r)}}
+O(k^2).\label{eq:2.34}
\end{equation}

Our task is now to study the existence of solutions $A(r)$ and $B(r)$ of
the two integral equations (\ref{eq:2.30}) and (\ref{eq:2.31}), and more
specifically, to study the behaviour of $A$ and $B$ for small~$r$.

In Appendix~B, we shall prove that for the general class of potentials,
$V(r)$, satisfying
\begin{eqnarray}
\mbox{A)}&\qquad&\int_0^{\infty}r'\,dr'\,|V(r')|\,(|\ln r'|+1)
<\infty\label{eq:2.35}\\
\noalign{\noindent \mbox{and}}
\mbox{B)}&\qquad&\int_a^{\infty}r'\,dr'\,|V(r')|\,(\ln
r')^2<\infty,\quad a>1,\label{eq:2.36}
\end{eqnarray}
the solutions $A(r)$ and $B(r)$ exist for all $r>0$, and furthermore,
near $r=0$ one has the behaviour
\begin{equation}
A(r)=[-gC_a(g)+o(1)]\ln r\label{eq:2.37}
\end{equation}
and
\begin{equation}
B(r)=[1-gC_b(g) +o(1)]\ln r.\label{eq:2.38}
\end{equation}
Here,
\begin{equation}
C_a(g)=\int_0^{\infty} r\,dr\,V(r)A(r)\label{eq:2.39}
\end{equation}
and
\begin{equation}
C_b(g)=\int_0^{\infty} r\,dr\,V(r)B(r).\label{eq:2.40}
\end{equation}
Both integrals for $C_a$ and $C_b$ are absolutely convergent since one
can easily show that, as $r\to \infty$, $\,A$ and $B$ have the bounds
\begin{equation}
|A(r)|<\mbox{Const.},\qquad |B(r)|<\hbox{Const.}\,|\ln r|,\label{eq:2.41}
\end{equation}
for $r>r_0>1$.  The convergence of Eqs.\ (\ref{eq:2.39}) and
(\ref{eq:2.40}) at $r=0$ is guaranteed by Eqs.\ (\ref{eq:2.35}),
(\ref{eq:2.37}), and (\ref{eq:2.38}).

Going back to Eq.\ (\ref{eq:2.34}), we write for the neighborhood of
$r\approx 0$:
\begin{equation}
-{g(k,r)\over \tilde f_{-}(k,r)}={(\pi/2)gC_a(g)\ln r +O(1)\over
{\displaystyle gC_a(g)\ln r\,\biggl(\ln k -\ln 2+\gamma
-i\,{\pi\over 2}\biggr) +[gC_b(g)-1]\ln r +O(1)}} +O(k^2).\label{eq:2.42}
\end{equation}
This result leads to
\begin{equation}
e^{i\delta_0(k)}\sin\delta_0(k)={\pi\over 2}\left[\raise
4pt \hbox{${\displaystyle {gC_a(g)\over
{\displaystyle gC_a(g)\biggl(\ln k -\ln 2+\gamma
-i\,{\pi\over 2}\biggr) +[gC_b(g)-1]}}}$}\right]+O(k^2).\label{eq:2.43}
\end{equation}

There are now two cases to consider, $\,C_a(g)\ne 0$ and $C_a(g)= 0$.
For $C_a(g)\ne 0$, we have the universal result as $k\to 0$,
\begin{equation}
\delta_0(k)={\pi\over 2\ln k}+O\biggl({1\over (\ln
k)^2}\biggr).\label{eq:2.44}
\end{equation}
One should note that $C_b(g)$ is finite. A somewhat stronger form of
(\ref{eq:2.44}) is that, as $k\to 0$,
\begin{equation}
e^{i\delta_0(k)}\sin \delta_0(k)={\pi\over 2\ln k-i\pi}+O\biggl({1\over (\ln
k)^{2,3}}\biggr),\label{eq:2.45}
\end{equation}
meaning that the real part of the first term is accurate to order $(\ln
k)^{-2}$ while the imaginary part is accurate to $(\ln k)^{-3}$.

The second case, $\,C_a(g)=0$, is clearly exceptional.  If $C_a(g)=0$ for
any interval $g_1<g<g_2$, then $V\equiv 0$.  For $V\not\equiv 0$,
$\,C_a(g)$ can only vanish for discrete values of~$g$.  In this case,
because of (\ref{eq:2.38}) and (\ref{eq:2.39}), $\,(1-gC_b)$ cannot
vanish.  Hence, it follows from (\ref{eq:2.42}) that, as $k\to 0$,
\begin{equation}
\delta_0(k)=O(k^2).\label{eq:2.46}
\end{equation}

Equation (\ref{eq:2.43}) also implies the uniform formula
\begin{equation}
\delta_0(k)={\xi\over\xi +1}\,{\pi\over 2\ln k -i\pi} +O\biggl({1\over
(\ln k)^{2,3}}\biggr)\label{eq:2.47}
\end{equation}
in the same sense as (\ref{eq:2.45}), where
\begin{equation}
\xi={{\displaystyle gC_a(g)\biggl(\ln k-i\,{\pi\over 2}\biggr)}\over
gC_b(g)-1}.\label{eq:2.48}
\end{equation}

\section{THRESHOLD BEHAVIOUR IN $(2+1)$ DIMENSIONS:\protect\\ THE FIELD
THEORETICAL CASE}
\setcounter{equation}{0}

We take as our starting point axiomatic local field theory with a minimum
non-zero mass.  There is then very little difference between \hbox{$2+1$}
and \hbox{$3+1$} dimensions.  In both cases, the on-shell scattering
amplitude depends on two variables.  The analyticity domain of the
scattering amplitude is obtained, in both cases, in two steps:\ \ i)
analytic continuation of the off-shell amplitude
\cite{seven},\cite{eleven} and ii) use of the
positivity of the absorptive part to enlarge the analyticity
domain \cite{twelve}.The partial wave expansion in the
\hbox{$(2+1)$}-dimensional case is given in terms of Chebyshev polynomials
and not Legendre polynomials.  Indeed, for the \hbox{$(2+1)$}-dimensional
case, we have
\begin{equation}
T(s,\cos\theta)=16\sum_{n=0}^{\infty}\,\epsilon_nf_n(s)\cos
n\theta.\label{eq:3.1}
\end{equation}
Here, $\,s$ is the square of the center-of-mass energy, and $\theta$ is the
scattering angle.  In the elastic region, $\,f_n(s)$ is related to the
phase shifts by
\begin{equation}
f_n(s)=\sqrt{s}\,e^{i\delta_n}\,\sin\delta_n.\label{eq:3.2}
\end{equation}
This and the factor of 8 in (\ref{eq:3.1}) are chosen to give
$T(s,\cos \theta)=-g+O(g^2)$ in $\phi_3^4$ perturbative field theory
with a $(g/4!)\phi^4$ interaction.

The absorptive part of $T$ is
\begin{equation}
A_s(s,\cos\theta)=16\sum_{n=0}^{\infty}\,\epsilon_n\cdot {\rm Im}\
f_n(s)\cos n\theta,\label{eq:3.3}
\end{equation}
with ${\rm Im}\ f_n(s)\ge 0$, from the unitarity condition.  From Eq.\
(\ref{eq:3.3}), it is easy to obtain
\begin{equation}
\biggl|\biggl({d\over d\cos\theta}\biggr)^nA_s(s,\cos\theta)\biggr| \le
\biggl({d\over
d\cos\theta}\biggr)^nA_s(s,\cos\theta)\biggr|_{\cos\theta=1};\quad
s\ge 4m^2,\label{eq:3.4}
\end{equation}
for all $\theta$ such that $-1\le \cos\theta\le +1$.  This last
inequality is precisely what made the enlargement of the analyticity
domain in the $3+1$ case possible \cite{twelve}.Therefore, one gets
the same enlargement in $2+1$ dimensions.

For simplicity, we consider a case with the kinematics and symmetry of
pion-pion scattering although our results are much more general.  We
use the Mandelstam variables
\begin{eqnarray}
s&=&4(k^2+m^2),\nonumber\\
t&=&2k^2(\cos\theta -1),\nonumber\\
u&=&4m^2-s-t.\label{eq:3.5}
\end{eqnarray}

For any fixed $t$, $\,|t|<4m^2$, $\,T(s,t)$ is analytic in the doubly
cut $s$-plane with cuts along
\begin{eqnarray}
s&=&4m^2+\lambda,\nonumber\\
u&=&4m^2+\lambda;\quad \lambda>0.\label{eq:3.6}
\end{eqnarray}
For fixed $s$, the absorptive part, $\,A_s(s,\cos\theta)$, is analytic
inside an ellipse in the $\cos\theta$-plane, which is an enlargement of
the Lehmann ellipse \cite{thirteen}.The foci are at $\cos\theta=\pm 1$
and the right extremity is at $\cos\theta=1+4m^2/2k^2$.

The partial wave amplitudes, $\,f_n(s)$, are defined as
\begin{equation}
f_n(s)={1\over 16\pi}\,\int_{-1}^{+1}\,T(s,\cos\theta)\,\cos
n\theta\,{d(\cos\theta)\over\sin\theta}.\label{eq:3.7}
\end{equation}
The $f_n$'s are analytic in a region that contains
\begin{equation}
|s-4m^2|<4m^2,\label{eq:3.8}
\end{equation}
excluding a cut along $4m^2\le s\le 8m^2$.  A major difference with the
$(3+1)$-dimensional case is the kinematical factor $\sqrt{s}$ which
comes from unitarity as explicitly shown in Eq.\ (\ref{eq:3.2}), a
point clarified with the help of R. Stora \cite{fourteen}.

Thus, the unitarity condition in $2+1$ dimensions is
\begin{equation}
{\rm Im}\ f_n(s)\ge {1\over\sqrt{s}}\,|f_n(s)|^2,\quad \forall\ 
s>4m^2.\label{eq:3.9}
\end{equation}
In the elastic region, $\,4m^2\le s<16m^2$,
\begin{equation}
{\rm Im}\ f_n(s)={1\over\sqrt{s}}\,|f_n(s)|^2.\label{eq:3.10}
\end{equation}

This slightly changed form of the unitarity condition given in Eq.\
(\ref{eq:3.9}), gives a different Froissart bound \cite{twelve},\cite{fifteen} in the $(2+1)$
case.  The number of partial waves effectively contributing to the
scattering amplitude is still bounded by
\begin{equation}
L=C\sqrt{s}\,\ln s,\label{eq:3.11}
\end{equation}
for large~$s$.  However, the Froissart bound in $2+1$ dimensions is
\begin{equation}
|F(s,\cos\theta)|<Cs\,\ln s,\quad -1\le\cos\theta\le +1.\label{eq:3.12}
\end{equation}
This is instead of the $s\ln^2 s$ in the $3+1$ case.  The number of
subtractions in the dispersion relations, for $|t|<4m^2$, is still at
most 2, as in the $3+1$ case \cite{sixteen}.

The general properties outlined so far are sufficient to determine the
singularity of $f_n(s)$ at $k=0$.  For simplicity, we restrict
ourselves to the $S$-wave case, although our method applies to the
higher waves.  It is convenient to change variables and define
\begin{equation}
f_0(s)=F_0(k).\label{eq:3.13}
\end{equation}
We also set the mass $m=1$.  In the variable $k$, the analyticity
domain of $F_0(k)$ contains the half circle $\Gamma$,
\begin{equation}
\Gamma:\quad \{|k|<1,\quad \mbox{and}\quad {\rm Im}\
k>0\}.\label{eq:3.14}
\end{equation}

A very important property of $T(s,t)$ is the reality property:\ \ $T$ is
real for $s<4$, $\,t<4$, $\,u<4$.  From this property, it follows that
$f_0(s)$ is real for $0<s<4$, and hence $F_0(k)$ is real for
$k=i\kappa$, $\,0<\kappa<1$.  By Schwarz's reflection principle, for
$k\in \Gamma$, we have
\begin{equation}
F_0(k)=F_0^*(-k^*).\label{eq:3.15}
\end{equation}

The unitarity condition, Eq.\ (\ref{eq:3.10}), can be written in a form
suitable for analytic continuation.  With initially $k=k^*$, we write
\begin{equation}
F_0(k)-F_0^*(k^*)={2i\over\sqrt{s}}\,F_0(k)F_0^*(k^*).\label{eq:3.16}
\end{equation}
This gives
\begin{equation}
F_0(k)={F_0^*(k^*)\over {\displaystyle
1-{2i\over\sqrt{s}}\,F_0^*(k^*)}},\label{eq:3.17}
\end{equation}
and defines a function analytic in the second sheet.  This function
will be the continuation to the semicircle, $\,|k|< 1$, $\,{\rm Im}\
k<0$, through the line $0<k<1$.  The only thing to prevent that would
be an accumulation of zeros of $[1-(2i/\sqrt{s})F_0^*(k^*)]$ along this
line, giving a natural boundary.  There is nothing in the general
axioms to prevent that \cite{seventeen}.However, it is sufficient to
assume that $F_0(k)$ is continuous on $0<k<1$ in order to avoid this
catastrophe.  We thus get the continuation of $F_0(k)$ to the second
sheet \cite{eighteen}, which, using the reality condition (\ref{eq:3.15}), can
be written as
\begin{equation}
F_0(k)={F_0(-k)\over {\displaystyle 1-{2i\over
\sqrt{s}}\,F_0(-k)}}.\label{eq:3.18}
\end{equation}
Hence, $\,F_0(k)$ is meromorphic for $|k|<1$, outside the origin.

Let us introduce $G_0(k)$ as
\begin{equation}
G_0(k)={1\over F_0(k)}.\label{eq:3.19}
\end{equation}
We get
\begin{equation}
G_0(k)=G_0(-k)-{2i\over\sqrt{s}}.\label{eq:3.20}
\end{equation}
Next, we define $H_0(k)$ as
\begin{equation}
H_0(k)\equiv G_0(k)-{2\over\pi\sqrt{s}}\,\biggl(\ln k-i\,{\pi\over
2}\biggr).\label{eq:3.21}
\end{equation}
$H_0(k)$ is again real for $k=i\kappa$, $\,0<\kappa<1$.  Using Eq.\
(\ref{eq:3.21}), we get
\begin{equation}
H_0(k)=H_0(-k).\label{eq:3.22}
\end{equation}
$H_0$ is therefore an even function of $k$, i.e.,
\begin{equation}
H_0(k)\equiv K_0(k^2).\label{eq:3.23}
\end{equation}
$K_0(k^2)$ is a meromorphic function of $k^2$, and the $S$-wave
amplitude can be written as
\begin{equation}
F_0(k)={1\over{\displaystyle K_0(k^2)+{2\over \pi\sqrt{s}}\,\biggl(\ln
k -i\,{\pi\over 2}\biggr)}}.\label{eq:3.24}
\end{equation}
If $K_0(k^2)$ has no pole at the origin, the $\ln k$ dominates the
denominator as $k\to 0$, and we get
\begin{equation}
F_0(k)\simeq {\pi\over 2}\,\sqrt{s}\,\biggl({1\over\ln
k}\biggr).\label{eq:3.25}
\end{equation}   
The phase shift then behaves as
\begin{equation}
\delta_0(k)\simeq {\pi\over 2\ln k},\label{eq:3.26}
\end{equation}
which is precisely the behaviour obtained in the potential case.  As in
the potential case, the existence of a pole of $K_0(k^2)$ at $k^2=0$
cannot be excluded.

The derivation we presented above also applies to higher waves, but it
can be proved that what is hopefully an exception for $n=0$ turns out to
be the rule for $n\ge 1$.  $K_n(k^2)$ has a pole, and we shall show in a
future publication that $\delta_n\sim k^{2n}$ for $n\ge 1$.

For the restricted class of potentials such that
\begin{displaymath}
\int_0^{\infty}
r\,dr\,\Bigl|1+|\ln r|\Bigr|\,\,|V(r)|\exp \mu r<\infty,
\end{displaymath}
the derivation of the dispersion relations for $|t|<\mu^2$ obtained first
by one of us \cite{nineteen} in the $(3+1)$ case also holds in $2+1$ dimensions.  It
implies that the partial wave amplitude is analytic in $|k|<\mu/2$, $\,{\rm
Im}\ k>0$, and therefore the derivation presented in this section applies
also to this potential case.

Equation (\ref{eq:3.17}) was also obtained by Bros and Iagolnitzer in
ref. \cite{six},  in a more general but less
elementary approach based on a postulated analyticity of the $S$-matrix
which, however, becomes axiomatic by using, as we said in the
introduction, ref. \cite{seven}. 
These authors emphasize the Riemann sheet structure at the threshold
rather than the actual behaviour of the physical scattering amplitude.

\section{PERTURBATION THEORY FOR $\phi_3^4$}
\setcounter{equation}{0}

It is of importance to compare our result with perturbation theory.  We
are fortunate that in \hbox{$(2+1)$} dimensions we have a rigorously defined
super-renormalizable theory \cite{one} with a mass gap, namely, $\,\phi_3^4$.

Taking 
\begin{displaymath}
{\cal L}_{\rm int}(\phi)={-g\over 4!} : \phi^4(x) :
\end{displaymath}
we obtain up to order $g^2$ for $T(p_1,p_2;-p_3,-p_4)$
\begin{equation}
T(s,t)=-g+g^2[f(s)+f(t)+f(u)]+O(g^3),\label{eq:4.1}
\end{equation}
where $f(s)$, $\,s=(p_1+p_2)^2$, is given by the Feynman diagram shown in
Figure~1,
\begin{equation}
f(s)=\biggl({-i\over 2}\biggr)\int\,{d^3k\over (2\pi)^3}\,{1\over
(k^2-\mu^2+i\epsilon)((p_1+p_2-k)^2-\mu^2+i\epsilon)}.\label{eq:4.2}
\end{equation}
The factor $({1\over2})$ is for identical outgoing particles, and the
$(-i)$ follows from $S=1+iT$, $\,S$ being the $S$-matrix.

This last integral can be easily evaluated in the Euclidean region,
$\,s<4\mu^2$, by carrying out a Wick rotation, and the result is
\begin{equation}
f(s)=-{1\over 16\pi\sqrt{s}}\,\ln\biggl({2\mu-\sqrt{s}\over 2\mu+\sqrt{s}}
\biggr),\quad 0<s<4\mu^2.\label{eq:4.3}
\end{equation}

The normalization of $T$ is chosen such that elastic unitarity is given by
\begin{equation}
{1\over 2i}\,(T-T^*)={1\over 16\sqrt{s}}\,\int_0^{2\pi}\,{d\theta\over
2\pi}\,|T(s,\theta)|^2,\quad 4\mu^2\le s<16\mu^2.\label{eq:4.4}
\end{equation}
The partial wave expansion is then
\begin{equation}
T(s,\theta)=16\sqrt{s}\,\sum_n\,\epsilon_n\,\cos
n\theta\,\,e^{i\delta_n}\sin\delta_n.\label{eq:4.5}
\end{equation}

As $s\to 4\mu^2$, $\,k\to 0$, then for physical $\theta$, $\,t\to 0$,
$\,u\to 0$, and the leading log term comes from (\ref{eq:4.3}), since
$f(0)$ is finite.

We get for $k\to 0$
\begin{equation}
T=-g-{g^2\over 32\pi\mu}\,\ln
{k^2\over\mu^2}+O(1)g^2+O(g^3).\label{eq:4.6}
\end{equation}
The first thing to notice is that at order $g^2$, $\,T$ diverges as $k\to
0$.  This is just the opposite of the full result we obtained in the
previous section where $T\to 0$ as $k\to 0$.

In third order, the leading $\ln k$ behaviour comes from the two-bubble
diagram shown in Figure~2.  The triangle diagram in Figure~3 is only of
order $(\ln k)$.  We conjecture that this continues in higher orders, and
the leading $(\ln k)$ approximation is given by
\begin{equation}
T\simeq -g\,\sum_{n=0}^{\infty}\,\biggl({g\ln (k/\mu)\over
16\pi\mu}\biggr)^n,\quad k\to 0.\label{eq:4.7}
\end{equation}
This sum is divergent for $k<\mu\exp(-16\pi\mu/g)$.  Thus the present
perturbation calculation does not give a meaningful result.  If we ignore
this divergence and sum the geometric series formally, the result is
\begin{equation}
T\simeq -g\left\{\raise
4pt \hbox{${\displaystyle {1\over {\displaystyle 1-\biggl({g\ln (k/\mu)\over
16\pi\mu}\biggr)}}}$}\right\}.\label{eq:4.8}
\end{equation}
However, as $k\to 0$, $\,s\to 4\mu^2$,
\begin{equation}
{T\over 16\sqrt{s}}\simeq e^{i\delta_0}\sin\delta_0.\label{eq:4.9}
\end{equation}
We thus recover the potential scattering result as $k\to 0$,
\begin{equation}
e^{i\delta_0(k)}\sin \delta_0(k)\sim {-g\over 32\mu}\,\left( \raise
4pt \hbox{${\displaystyle {1\over {\displaystyle -{g\ln(k/\mu)\over
16\pi\mu}}}}$}\right)\label{eq:4.10}
\end{equation}
and
\begin{equation}
\delta_0(k)\sim {\pi\over 2}\,{1\over
\ln (k/\mu)},\quad k\to
0.\label{eq:4.11}
\end{equation}

$\phi_3^4$ is a well-defined theory, both perturbatively and
non-perturbatively, and it is clear from our results that as $k\to 0$
perturbation theory gives the wrong answer.  It is perhaps interesting to
note that $\phi_3^4$ is asymptotically free.  If our conjecture on the
higher-order ($\ln k$) behaviour is correct, then this would be the first
completely rigorous demonstration of how perturbation theory order by
order could be extremely misleading. Such a resummation is also present
in ref. \cite{six}.

\section{REMARKS AND CONCLUSIONS}
\setcounter{equation}{0}

We close this paper with three significant remarks.

i) The power of elastic unitarity together with analyticity is clearly
demonstrated by the following remark stressed to us by
Porrati \cite{twenty}. Once we are given a phase-shift behaviour such
that
\begin{equation}
a_0(k)=e^{i\delta_0(k)}\sin\delta_0(k)={c\over {\displaystyle \ln k-
i\,{\pi\over 2}}} +O\left({1\over (\ln k)^{1+\epsilon}}\right),\quad k\to
0,\label{eq:5.1}
\end{equation}
then unitarity alone fixes $c$ to be $c=\pi/2$, since
\begin{equation}
a_0^*(k)=a_0(-k)={c\over {\displaystyle \ln |k| -i\,{\pi\over 2}}}+O\left(
{1\over (\ln \mbox{$-\mskip-2mu k$})^{1+\epsilon}}\right).\label{eq:5.2}
\end{equation}
The factors $(i\pi/2)$ are necessary to keep $a_0(k)$ real for $k$ purely
imaginary and ${\rm Im}\ k>0$.  Hence we get
\begin{equation}
{\rm Im}\ a_0(k)={\pi\over 2}\,{c\over (\ln k)^2}+O\left( {1\over
(\ln k)^{2+\epsilon}} \right).\label{eq:5.3}
\end{equation}
From ${\rm Im}\ a_0=|a_0|^2$, we obtain, when $c\ne 0$,
\begin{equation}
c=\pi/2.\label{eq:5.4}
\end{equation}
It should be pointed out, however, that this argument requires
analyticity in $k$ in a semicircle in ${\rm Im}\ k>0$, and hence only
applies to exponentially decreasing potentials.

ii) In one dimension, the simplest potential is the $\delta$-function
potential.  In two or three dimensions, the corresponding simplest
potential is the so-called point interaction, which is the same as the
Fermi pseudopotential.  There is a large literature on the Fermi
pseudopotential.

Recently, Jackiw \cite{twentyone} obtained the phase shift $\delta_0(k)$ for the
point interaction in two dimensions.  Although this potential does not
belong to the class considered in Section 2, his result for $k\to 0$
agrees with that of ref. \cite{one} and ours; see Eq.\ (3.26) in his paper.  It
should be stressed, however, that our relativistic result holds for any
$2+1$ field theory with the standard analyticity and without zero-mass
particles; we are not restricted to $\phi_3^4$.

iii) In a $\phi^4$-type field theory, the renormalized coupling constant is
defined by the value of the $2\to 2$ scattering amplitude, $\,T(s,t,u)$,
evaluated at some Euclidean point $(s,t,u)< 4\mu^2$, often for
convenience taken to be the symmetric point $s=t=u=4\mu^2/3$.  In
$(3+1)$ dimensions, given the well-established analyticity and unitarity
properties of $T$, it has been shown in many papers \cite{twentytwo} that the
coupling constant is bounded.  Some of these
bounds are surprisingly strong.  In $\phi_3^4$, Glimm and Jaffe
\cite{one}
obtained bounds directly from constructive field theory, but their
results are weaker than what can be obtained from analyticity and
unitarity.

The general methods used in the papers cited in ref. \cite{twentytwo} for the
$(3+1)$ case can be easily modified to apply to $(2+1)$ dimensions. 
Only the kinematic factor outside the partial wave expansion is
different.  The results of this paper thus present us with a new and
significant challenge.  We have now a new piece of information on the
scattering amplitude which is exact.  Namely, we know that 
\begin{displaymath}
T(s,t,u)\,\ln {\sqrt{s-4\mu^2}\over 2\mu}\to
16\pi\mu\quad\mbox{as $s\to 4\mu^2$, $\,t\to 0$, $\,u\to 0$,} 
\end{displaymath}
i.e., at certain points on the
Mandelstam triangle.  Given the power of unitarity and
analyticity, we are quite confident that this new input will improve the
bounds on the coupling constant.  Only the magnitude of the improvement
is in question.  Work on this problem is in progress.

\vspace*{1cm}

\noindent{\bf ACKNOWLEDGEMENTS}

We thank our colleagues H. Grosse, H. Lehmann, M.
Porrati, and H.~C. Ren for their useful comments, and R. Stora for his
crucial help.  We also thank J. Bros for pointing out to us his previous
work.  Two of us, N.N.K. and T.T.W., are grateful to the CERN Theory
Division for its kind hospitality.  This work was supported in part by the
U.S.\ Department of Energy under Grant No.\ DE-FG02-91ER40651, Task B, and
under Grant No.\ DE-FG02-84ER40158.

\vfil
\eject


\setcounter{section}{0}
\setcounter{equation}{0}
\renewcommand{\thesection}{Appendix \Alph{section}.}
\renewcommand{\theequation}{A. \arabic{equation}}
\section{}

In this appendix, we study briefly the equation (\ref{eq:2.2}), together
with the Dirichlet boundary condition (\ref{eq:2.3}).  We start with the
free equation,
\begin{equation}
\biggl({d^2\over dr^2}+{1\over 4r^2}+k^2\biggr)u(k,r)=0,\label{eq:A1}
\end{equation}
with $u(k,0)=0$.  Because of the presence of the attractive singular
potential, $\,-1/(4r^2)$, one must be careful in the extension of the
differential operator, $\,-(d^2/dr^2)-(1/4r^2)$, to a self-adjoint operator
on $L^2(0,\infty)$.  This has been thoroughly studied in the
literature \cite{twentythree},\cite{twentyfour}. We quote the result here.  The two
independent fundamental solutions of (\ref{eq:A1}) are $\sqrt{r}\,J_0(kr)$
and $\sqrt{r}\,Y_0(kr)$.  Both vanish at the origin.  Every other
solution, being a linear combination of these two, also vanishes at
$r=0$.  Therefore, we are in the limit-circle case for the differential
operator with a Dirichlet boundary condition at $r=0$.  There exist an
infinite number of self-adjoint extensions of the symmetric differential
operator, depending on one (real) parameter.  Each self-adjoint extension
is defined by the amount of mixing of the two fundamental solutions. 
Among all these extensions, there exists a ``distinguished'' one, which
corresponds to taking the pure Bessel solution $\sqrt{r}\,J_0(kr)$.  These
generalized eigenfunctions are less singular, behaving like $\sqrt{r}$ at
the origin, as compared to the eigenfunctions of all other extensions,
which behave like $\sqrt{r}\,\ln r$ as $r\to 0$.  Moreover, it can be
shown that the ``distinguished'' extension corresponds to the Friedrichs
extension \cite{twentyfour},\cite{twentyfive}. But, for the physicist, the more
important fact is this:\ \ in all the other self-adjoint extensions, there
exists, besides the continuum, a negative energy eigenvalue.  In other
words, there exists always a real bound state with negative energy,
$\,E_0=k_0^2<0$ \cite{twentythree},\cite{twentyfour}.

The extension $H_{\lambda}$ is defined by taking the behaviour, as $r\to 0$,
\begin{equation}
u(r)\to \sqrt{r} +\lambda\sqrt{r}\,\ln r;\quad \lambda\
\mbox{real}.\label{eq:A2}
\end{equation}
It is then easy to check that if we define a solution such that
\begin{equation}
\sqrt{r}\,[J_0(kr)+Y_0(kr)]\tolim_{r\to 0} \sqrt{r}+\lambda\sqrt{r}\,\ln
r,\label{eq:A3}
\end{equation}
then an elementary calculation shows that, by setting $k=+i\kappa_0$, we get
\begin{equation}
\ln\kappa_0={1-\lambda(\gamma-\ln 2)\over\lambda},\label{eq:A4}
\end{equation}
where $\gamma$ is the Euler constant.  Thus, for any real $\lambda$,
$\lambda>0$, we have a bound state at $E=-\kappa_0^2(\lambda)$.

There is no such bound state in the ``distinguished'' extension.  However,
in this case we are just at the threshold of having a bound state.  More
precisely, in the ``distinguished'' extension, if we add to the free
Hamiltonian a purely attractive (negative) potential, no matter how weak it
happens to be, there appears a true bound state.  This fact is well
established in the literature using a variational argument.

As an aside here we give the upper bound of Set\^o \cite{twentysix} on the number
$N$ of bound states for dimension $\null=2$, and $l=0$.  This is the
2-dimensional version of the old Bargmann inequality for $d=3$.  The
Set\^o bound is
\begin{equation}
N_2^0\le 1+{{\displaystyle {1\over 2}\int_0^{\infty}dr\int_0^{\infty}
dr'\,\left|\,\ln {r\over r'}\,\right|\,V(r)\,V(r')}\over {\displaystyle
-\int_0^{\infty}r\,V(r)\,dr}},\label{eq:A5}
\end{equation}
where, given our assumptions on $V(r)$, all the integrals are finite.  The
fact that there is always a bound state, regardless of how weak an
attractive potential $V$ may be, is somehow reflected by the presence of 1
in the right-hand side of (\ref{eq:A5}).  This cannot be improved.

In any case, this last property of the ``distinguished'' extension of the
free differential operator to a self-adjoint operator {\em without a
bound state} is the most important criterion by which we must choose this
extension, and discard all others.  As physicists, we do not have the
freedom to start with a ``free Hamiltonian'' that binds a free particle. 
Mathematicians have this luxury.

We finally come to the equation (\ref{eq:2.2}) itself.  Starting from the
``distinguished'' extension of the free Hamiltonian, and adding to it a
potential $V$, does not alter the self-adjointness, provided $V$ is ``weak''
in the sense of Kato and others \cite{twentyfive},\cite{twentyseven}. 
The condition
defining this ``weak'' class is expressed precisely in the following
integrability condition on the potential:
\begin{equation}
\int_0^{\infty}r\,dr\,(1+|\ln r|)\,|V(r)| <\infty.\label{eq:A6}
\end{equation}
This ensures the semi-boundedness of the total Hamiltonian, and the
finiteness of the number of bound states.  Note that (\ref{eq:A6}) is
precisely the condition (\ref{eq:2.35}) which we had to use in
section~II.  We shall need it in Appendix~B to establish the existence and
study the properties of the solutions of the two integral equations
(\ref{eq:2.30}) and (\ref{eq:2.31}).

To conclude this appendix, let us point out that an extension different
from the ``distinguished'' one can be used to simulate a renormalized
delta-function interaction, as was done by Jackiw \cite{twentyone}.

\setcounter{equation}{0}
\renewcommand{\thesection}{Appendix \Alph{section}.}
\renewcommand{\theequation}{B. \arabic{equation}}
\section{}

In this appendix we study the integral equations (\ref{eq:2.30}) and
(\ref{eq:2.31}).  For the class of potentials satisfying Eqs.\
(\ref{eq:2.35}) and (\ref{eq:2.36}), we first prove that the solutions
$A(r)$ and $B(r)$ exist and are bounded, as $r\to\infty$, as in Eq.\
(\ref{eq:2.41}).  Next, we prove that the behaviour of $A(r)$ and $B(r)$ as
$r\to 0$ is given by Eqs.\ (\ref{eq:2.37}) and (\ref{eq:2.38}),
respectively.  We will only give the details for Eq.\ (\ref{eq:2.31}).  The
procedure for Eq.\ (\ref{eq:2.30}) is easier and very similar.

Our starting point is the integral equation
\begin{equation}
B(r)=\ln r + g\int_r^{\infty}r'\,dr'\,\biggl(\ln {r'\over r}\biggr)
V(r')B(r').\label{eq:B1}
\end{equation}
We can first consider the case $r'\ge r\ge 1$, where we have the inequality
\begin{equation}
0\le \ln {r'\over r} \le \ln r'.\label{eq:B2}
\end{equation}
Therefore, an upper bound $\bar B$ is obtained for $B$ by replacing the
integral equation (\ref{eq:B1}) by
\begin{equation}
\bar B(r)=\ln r +g\int_r^{\infty}r'\,dr'\,|V(r')|\,\ln r'\,\bar
B(r'),\quad r>1.\label{eq:B3}
\end{equation}
The solution of (\ref{eq:B3}) can be obtained by standard methods and is
given by
\begin{equation}
\bar B(r)=\biggl[\int_1^r\,{dt\over t}\,\exp\biggl(-g\int_t^{\infty}
u\,|V(u)|\,\ln u\,du\biggr) +C\biggr] \exp\biggl(g\int_r^{\infty}
t\,|V(t)|\,\ln t\,dt\biggr).\label{eq:B4}
\end{equation}
The constant $C$ is given by
\begin{equation}
C=\int_1^{\infty}\,{1\over r}\,\biggl[1-\exp\biggl(-g\int_r^{\infty}
t\,|V(t)|\,\ln t\,dt\biggr)\biggr]\,dr,\label{eq:B5}
\end{equation}
which is finite given (\ref{eq:2.35}).  Using this result in (\ref{eq:B4}),
we find that
\begin{equation}
\bar B(r)=[1 +o(1)]\ln r,\quad \mbox{as\ }r\to\infty.\label{eq:B6}
\end{equation}
This establishes the bound on $B(r)$ for $r\ge 1$,
\begin{equation}
|B(r)|\le C_1\ln r+D_1,\label{eq:B7}
\end{equation}
where $C_1$ and $D_1$ are positive constants depending on~$g$.

By the same technique, we arrive at similar conclusions for $A(r)$.  This
time, the bounding condition for $\bar A(r)$ is $\bar A(\infty)=1$.  We
obtain
\begin{equation}
\bar A(r)=1+o(1),\quad \mbox{as\ }r\to\infty\label{eq:B8}
\end{equation}
and
\begin{equation}
|A(r)|\le \bar A(r)\le D_2,\quad r\ge 1,\label{eq:B9}
\end{equation}
where $D_2$ is a positive constant.  

From these bounds one can easily get, as $r\to\infty$,
\begin{equation}
A(r)=1+o(1);\qquad B(r)=[1 +o(1)]\ln r.\label{eq:B10}
\end{equation}
It is important to note that for the first estimate we need only the
condition (\ref{eq:2.35}), whereas for the second we need
(\ref{eq:2.36}).

Finally, we consider the region $r<1$ for both $A(r)$ and $B(r)$.  The
case for $B(r)$ is more delicate (singular), and we treat it first.

We can write (\ref{eq:B1}) as
\begin{equation}
B(r)=\ln r+g\int_r^1r'\biggl(\ln {r'\over r}\biggr)\,V(r')B(r')\,dr'
+g\int_1^{\infty}r'\biggl(\ln {r'\over
r}\biggr)V(r')B(r')\,dr'.\label{eq:B11}
\end{equation}
In the second integral, since $r'\ge 1$ and $r<1$, we can use the bound
(\ref{eq:B7}) and get, using condition (\ref{eq:2.29}),
\begin{equation}
\left|\int_1^{\infty}r\,\ln {r'\over r}\,V(r')B(r')\,dr'\right|<C+D\biggl(
\ln {1\over r}\biggr),\label{eq:B12}
\end{equation}
where $C$ and $D$ are positive constants.  In the first integral, we have
\begin{equation}
\left|\,\ln {r'\over r}\,\right|\le |\ln r|,\quad r<r'\le
1.\label{eq:B13} 
\end{equation}

An upper bound, $\,\bar {\mskip-3mu\bar B} (r)$, for $B(r)$ in $r\le 1$ is
now obtained by substituting (\ref{eq:B12}) and \mbox{(\ref{eq:B13})} in
(\ref{eq:B11}).  We obtain the integral equation
\begin{equation}
\bar {\mskip-3mu\bar B} (r) = C_2+D_2\,|\ln r|+g\,|\ln r|\,\biggl[
\int_r^1 r'\,|V(r')|\,\bar {\mskip-3mu\bar B}
(r')\,dr'\biggr],\label{eq:B14}
\end{equation}
with some positive constants $C_2$ and $D_2$.  

The solution of (\ref{eq:B14}) can be obtained by elementary methods.  It
is
\begin{eqnarray}
\bar {\mskip-3mu\bar B}(r)&=&Z(r)\,g\,|\ln r|\biggl[C_3+\int_r^1
r'|V(r')|[C_2 +D_2\,|\ln r'|] \Bigl(Z^{-1}(r')\Bigr)\,dr'\nonumber\\
&&\mbox{}+C_2+D_2\,|\ln r|,\label{eq:B15}
\end{eqnarray}
where
\begin{equation} 
Z(r)=\exp\biggl[
\int_r^1 dr'\,gr'\,|\ln r'|\,|V(r')|\biggr].\label{eq:B16}
\end{equation}
Noting that $Z(r)$ is bounded for $0\le r\le 1$, from the condition
(\ref{eq:2.35}), we get
\begin{equation}
|B(r)|\le \,\bar {\mskip-3mu\bar B} (r)<\lambda+\mu\,|\ln
r|.\label{eq:B17}
\end{equation}

In the same way, we can analyze the integral equation (\ref{eq:2.30})
for $A(r)$.  We again find that, for $r\to 0$,
\begin{equation}
|A(r)|\le \lambda_1\,|\ln r| +\mu_1.\label{eq:B18}
\end{equation}

Using these two bounds, we can now prove the asymptotic estimates
(\ref{eq:2.37}) and (\ref{eq:2.38}).  From Eq.\ (\ref{eq:2.30}), we get,
as $r\to0$,
\begin{equation}
A(r)=-gC_a\,\ln r+g\int_r^{\infty} r'\,\ln r'\,V(r')A(r')\,dr' +
1.\label{eq:B19}
\end{equation}
This can be written as
\begin{equation}
A(r)=-gC_a\,\ln r+g\int_r^1 r'\,\ln r'\,V(r')A(r')\,dr'
+O(1).\label{eq:B20}
\end{equation}
The integral in (\ref{eq:B20}) could diverge as $r\to 0$.  However, setting
\begin{equation}
I(r)=g\int_r^1r'\,\ln r'\,V(r')A(r')\,dr',\label{eq:B21}
\end{equation}
and using (\ref{eq:B18}), we get
\begin{eqnarray}
|I(r)|&<&g\lambda_1\int_r^1r'\,|\ln r'|^2|V(r')|\,dr'
+ g\mu_1\int_r^1r'\,|\ln r'|\,|V(r')|\,dr'\nonumber\\
&<&g\lambda_1\int_r^1r'\,|\ln r'|^2|V(r')|\,dr' + O(1).\label{eq:B22}
\end{eqnarray}
Next we define
\begin{equation}F(r)\equiv r^2\,|\ln r|^2\,|V(r)|.\label{eq:B23}
\end{equation}

From the condition (\ref{eq:2.35}), we have
\begin{equation}
\int_0^1dr'\,r'\,|\ln r'|\,V(r)=\int_0^1\,{dr'\over r'\,|\ln r'|}\cdot
F(r') < \mbox{const.}\label{eq:B24}
\end{equation}
This implies that $F(r)\to 0$ as $r\to 0$.  From Eqs.\ (\ref{eq:B22}) and
(\ref{eq:B23}), we get
\begin{equation}
|I(r)|\le g\lambda_1\int_r^1\,{dr'\over r'}\,F(r') + O(1),\label{eq:B25}
\end{equation}
and, hence, since $F(r')$ vanishes as $r'\to 0$,
\begin{equation}
|I(r)|= |\ln r|o(1).\label{eq:B26}
\end{equation}
This establishes Eq.\ (\ref{eq:2.37}).  For Eq.\ (\ref{eq:2.38}), the
derivation is similar.

It is important to notice that, if $A(r)/\ln r\to 0$ as $r\to 0$, then
$B(r)/\ln r$ cannot approach zero as $r\to 0$.   This is because $A$ and
$B$ are solutions of the {\em same} differential equation,
\begin{displaymath}
{d\over dr}\biggl(r\,{dX\over dr}\biggr)=-grV(r)X(r),
\end{displaymath}
and are thus linearly independent.

\vfill\eject

\begin{figure}
\epsfig{figure=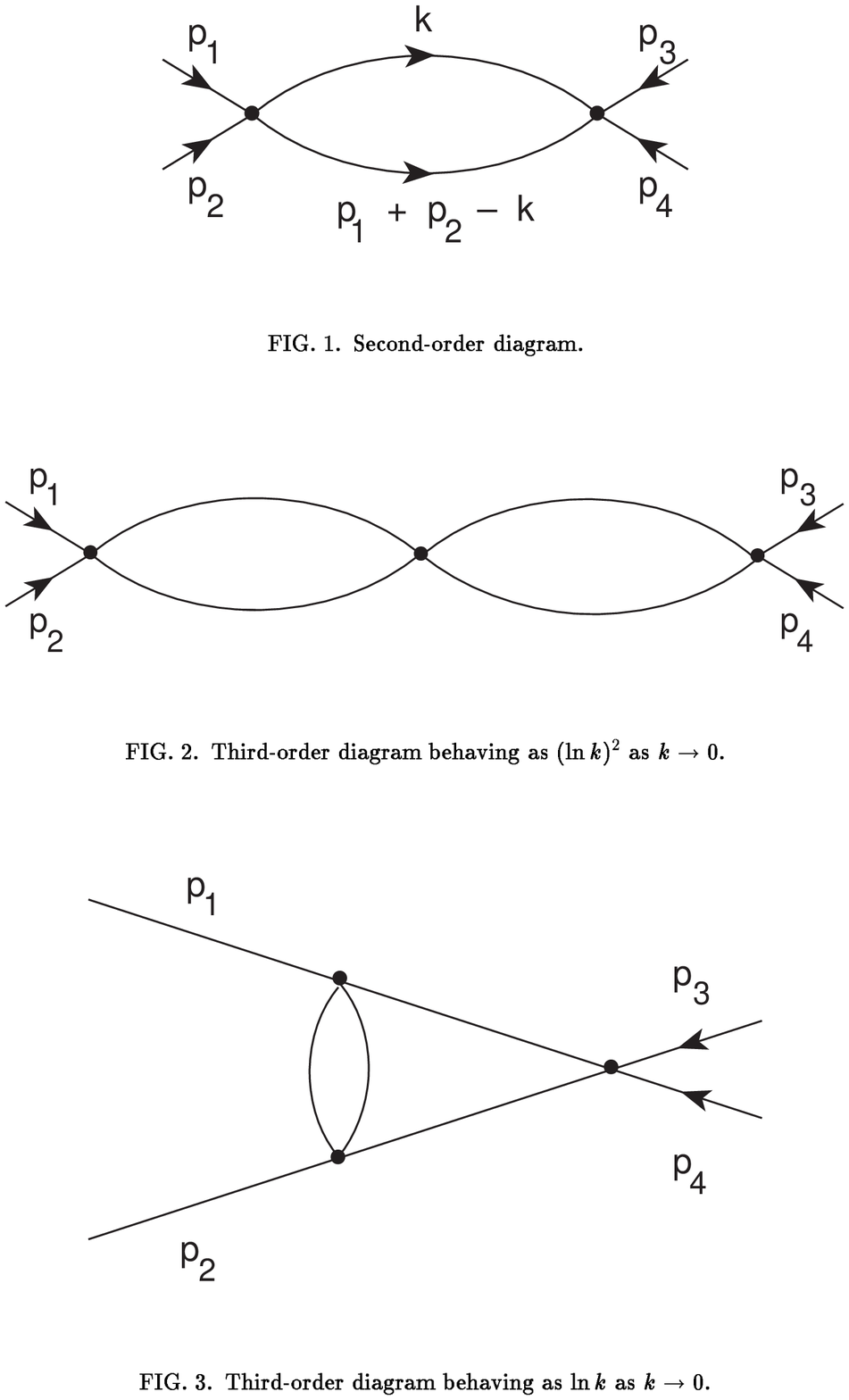,width=18cm}
\end{figure}
\end{document}